\documentclass[runningheads]{llncs}
\usepackage{graphicx}
\usepackage{mathrsfs} % Provides a \mathscr command
\usepackage{amssymb}
\usepackage{bm}
\begin{document}
\title{A vector logistic dynamical approach to epidemic evolution on interacting social-contact and production-capacity graphs.\thanks{Advances in Production Management Systems - APMS 2021 conference paper}
%, and its proxy economic cost assessment. 
}
\titlerunning{Vector logistic SIS dynamics on a coupled social-production graph}

\author{Jan B. Broekaert\inst{1}\orcidID{0000-0002-4039-8514} \and \\
Davide La Torre\inst{1}\orcidID{0000-0003-2776-0037} }
\authorrunning{J.B. Broekaert and D. La Torre}
\institute{AI Institute for Business,
SKEMA Business School,  Sofia Antipolis, France\\
\email{jan.broekaert@skema.edu, davide.latorre@skema.edu}\\
\url{https://www.skema.edu/faculty-and-research/artificial-intelligence}
%\and ABC Institute, Rupert-Karls-University Heidelberg, Heidelberg, Germany\\
}

\maketitle              % typeset the header of the contribution
\begin{abstract}
Population inhomogeneity, in the variation of the individual social contact networks and the individual infectious-recovery rates, renders the dynamics of infectious disease spreading uncertain.  As a consequence the overlaying economical production network with its proper collaboration components is to extent impacted unpredictably. Our model proposes a \emph{vector logistic} dynamical approach to SIS dynamics in a social contact network  interacting with its economic capacity network.    The probabilistic interpretation of the graph state in the  vector logistic description provides a method to assess the effect of mean and variance of the infected on the production capacity and allows the strategic planning of social connectivity regulation.  The impact of the epidemic mean effects and fluctuations on the production capacity is assessed according \emph{cumulative}, \emph{majority} and \emph{fragility} proxy measures.
\keywords{SIS dynamics \and vector logistic equation \and social graph \and production capacity.}
\end{abstract}

\section{Context and rationale}
% \section{On compartmental models - summary}
Compartmental epidemic models,  starting with Kermack and McKendrick \cite{KermackMcKendrick1927} and various elaborations reviewed by Hethcote \cite{Hethcote2000},  provide a simplified description of the epidemic evolution through transitions between a number of categories in a population, mainly the  {Susceptible},  {Infected},  and {Receptive} (`Recovered' or `Removed')  - and in further model extensions, the Exposed (latency of onset), the {Deceased} (change of population size)  and the  Maternal (immunity protection from birth), see e.g. \cite{Choisy2007}.   The aspects of socio-spatial distribution of individuals -  in terms of inhomogeneity of both the infection-recovery rates and the connectivity of each individual , and the \emph{stochastic} nature of transition events require an extension of the basic compartmental approach.  To encompass the effect of socio-spatial structure of the population in the endemic progression, and to cover the resulting fluctuations,  complex graph topologies have been implemented \cite{Keeling2005,Zhou2006},  and reformulated as the bond percolation problem \cite{Newman2002}.
Specific graph topologies have been related to epidemic extinction time by Ganesh et al. \cite{GaneshEtAl2005} and the resilience of epidemics related to the diameter of the underlying network (e.g. in the network structures of Facebook, the Internet, Social networks)  by Lu et al. \cite{LuEtAl2017}\\ %  and Erd\H{o}s–R\'enyi random  networks and scale-free networks
Non-deterministic epidemic models including the effects of fluctuations are based on stochastic diffusion equations,  by approximating the continuous time Markov chain model \cite{Allen2007},  by matching as an epidemic model with multiple hosts \cite{McCormack2006}, or  by parameter perturbation  \cite{Gray2011}.\\
Instead, our model develops a description of the epidemic dynamics immediately at the level of intrinsic infection probability, similar to probabilistic  Markov or `quantum-like' system descriptions, e.g. \cite{BusemeyerBruza2012,WangEtAl2013,Broekaert2020}.
In our model  we describe the interaction of two networks: the social contact network, as the graph union of all individual `ego' contact networks,   and the production capacity network as the graph union of all production clusters. The  nodes spanning both encompassing networks, $\mathscr{G}_A(V,E_A)$ and $\mathscr{G}_B(V,E_B)$, are the individuals of the considered population (filtered for professional activity in the production capacity graph).\\
We must take care to distinguish the concept of sub graphs - the connected components which do not share any edge with other such components, either in $\mathscr{G}_A(V,E_A)$ or $\mathscr{G}_B(V,E_B)$, and the interacting graphs which are (two) separate implementations - or layers - of functional relations on the same population. 
While the existence of connected components, or sub graphs,  in the social contact network $\mathscr{G}_A(V,E_A)$ has an effect on the dynamics of infectious disease spreading, and can be indirectly influenced by regulated restriction on the degree of social connectivity, such an intervention is not applicable in the economic capacity network.
The connected components of the production capacity graph remain fixed over time, since these components represent the economic capacity units of individual businesses, enterprises or service systems. It has been shown that, for interdependent networks failures in one network can percolate in another network on which its optimal performance depends \cite{Buldyrev2010}. In this manner, the interaction of the social graph with the production graph will allow an assessment of the production capacity attrition and will allow an analysis for the possible planning of regulatory intervention in the social contact network.\\
 Finally as we have shortly mentioned earlier, in our approach each node of the graph is characterised by its probability of being infected over time, instead of attributing to each node a binary status of ``infected'' or ``not-infected" at each instance of time. The evolution of the node infection probabilities is determined by the N-dimensional \emph{vector logistic} equation. A similar probabilistic infection approach on a graph was proposed by Wang et al. \cite{WangEtAl2003}, but which applied a Markov-like  dynamics (idem, Eq.13) instead. The usage of the vector logistic equation allows i) to regain the limit of the classic scalar logistic equation for SIS dynamics when the social graph nears the complete graph, $\mathscr{G}_A(V,E_A) = \mathscr{K}_N$ with large $N$, and ii) a probabilistic interpretation of the  graph state vectors of the nodes, $\mathbf{Y}$, in the unit $N$-hypercube $[0,1]^N$. This approach hence allows the expression of any infection related  expectation quantity $\langle \mathbf{f} \rangle_t$ = $E_t(\mathbf{f})$ = $\sum_i Y_i(t) f_i$.

%\citet{MassaroEtAl2018}

\section{Probabilistic SIS-dynamic on social contact graphs}
In the compartmentalised SIS-model the dynamics of  the infected fraction $i$, is determined by the recovery rate  $\delta$ over the infected fraction, and the infection rate  $\beta$ on the product of the susceptible, $s$, and infected, $i$, fractions: 
%\begin{eqnarray} 
%\dot i  &=&   - \delta i   + \beta   s i , \nonumber \\
%\dot s  &=&   \delta i   -  \beta   s i , \nonumber 
%\end{eqnarray}
% This results in the classical SIS logistic equation for the infected ratio $i=i(t)$ of the population, 
\begin{eqnarray} 
\dot i  &=&   - \delta i   + \beta   (1-i) i, \label{eq:SIS} 
 \end{eqnarray}
 where $1-i(t) = s(t)$. Two stationary solutions can occur $i_1^*=0$ and $i_2^* = 1-\delta/\beta$ (the latter when $\delta < \beta$). In our model, the possible interaction between the individuals - represented by nodes -  is controlled by the adjacency matrix $A$ of the graph $\mathcal{G}_A(V,E_A)$, $\vert V \vert = N$, $E_A \subset V^2$, representing the social contact network.
While the specific realisation of the adjacency matrix in the true social contact network remains unknown, a number of parameters can be estimated or assumed \cite{Barrett2009}. Some of its properties like the average degree can be regulated as an optimisation parameter, e.g. corresponding with the restricted number of  contacts that are allowed in a personal `social bubble' or `support circle'.

In order to retain a detailed description at the level of individual agents and to assess fluctuations over the network, a probability based infection-recovery model is constructed on a network. In this approach, a probability $Y_i$ of being infected is attributed  to each node $i$.
%\footnote{ In over-simplified version the time evolution of the infection probability of each node would follow a Markov dynamics.
%Under these simplified assumptions 
%\begin{equation} 
%\dot Y_i =   - \delta Y_i + \beta \sum_j A_{ij} Y_j 
%\end{equation}
%On the graph with adjacency matrix $A$, the Markov process evolution states
%or in matrix notation:
%\begin{equation} 
%\dot Y  =   - \delta Y  + \beta  A  Y 
%\end{equation}
%While this relates the change of $Y_i$ to $Y_j$ through their contact factor $A_{ij}$, this interaction expression does not distinguish between the state of $Y_i$, i.e. being infected or susceptible.}
Conform to the  interaction effect  by `contact', we express the  exposure of a node $i$  by the product of its proper receptive capacity $1-Y_i$ and the infective capacity  $Y_j$ of an adjacent node $j$, i.e. $A_{ij}=1$, weighted by infection rate $\beta_j$   and moderated by a normalisation factor  of the inverse of the node's degree $d_i^{-1}$.  
\begin{eqnarray} 
\dot Y_i =   - \delta_i Y_i + d_i^{-1} (1 - Y_i)\sum_j A_{ij}  \beta_j Y_j  \label{eq:vectorlogisticequationbycomp}
\end{eqnarray}
The dynamical equation of the graph state  vectors  is written in vector notation by  using $ \bm{\delta}$ both for the variable recovery rate vector and likewise $\bm{\beta}$   for the variable infection rate vector and for the state vector $\mathbf{Y}$. We further need the Hadamard product symbol, $\circ$, to express the \emph{elementwise} multiplication of factors:
\begin{eqnarray} 
\dot \mathbf{Y} =   -  \boldsymbol{\delta} \circ \mathbf{Y}  +  (  \mathbf{1}     -  \mathbf{Y} )    \circ  \mathbf{d}^{-1}   \circ A\ ( \boldsymbol{\beta}   \circ   \mathbf{Y} ) \label{eq:vectorlogistic}
\end{eqnarray}
  this notation requires that for an isolated node the apparent division `0/0' occurring in $A ( \boldsymbol{\beta}   \circ   \mathbf{Y} ) /\mathbf{d}$  is effectively set equal to 0. \\
This vector differential equation  differs from (the linear form of) the generalized Lotka-Volterra equation by a term proportional to $A  \mathbf{Y} $, and from the Replicator equation by  its additionally lacking a third order term $\mathbf{Y} \circ ( \mathbf{Y}^T A  \mathbf{Y} )$, and having a first-order term in $\mathbf{Y} $ instead.  Essentially the equation differs from these two typical dynamical systems by the first order derivative of the state {\bf not} being a Hadamard product with the state itself. 
In order to attribute a probabilistic interpretation to the magnitudes $Y_i$,  two observations are made,
\begin{itemize}
\item when $Y_i = 1$, the component $Y_i$  decays over time at rate $\delta_i$, 
\item when $Y_i =0 $, the component $Y_i$  grows at rate $1/d_i \, \mathbf{A}_{ij} ( \boldsymbol{\beta}   \circ   \mathbf{Y} )^j$, which is non-negative. 
\end{itemize}
With an initial state $ 0 \leq \mathbf{Y}_0 \leq 1$ at $t=0$, the component values of $\mathbf{Y}_t$ remain  contained in the $[0,1]$ range and  hence can be considered as event probabilities (for infection) assigned to the respective nodes  of the graph $\mathcal{G}(V,E)$. The state space of the vectors $\mathbf{Y}$ is the unit N-hypercube $[0,1]^N$, allowing each node an infection probability between 0 and 1. 
We recall that in contrast, in the Replicator system the corresponding state vector $\mathbf{Y}$ would remain on the simplex, $\sum_{i=1}^N Y_i = 1$, see e.g. Ohtsuki et al. \cite{Ohtsuki2006}, and in the case of the generalized Lotka-Volterra equation the solution is unconstrained $Y \in {\mathbb{R}^+}^N$.

 It can be easily shown that the vector logistic system reduces to the standard compartmentalised SIS equation when the graph is complete $\mathcal{G}(V,E) = K_N$ (all nodes have grade $N-1$), and the recovery and infection rates are considered constant over the graph. With $i= \frac{1}{N}\sum_j Y_j$ and $A \mathbf{Y} =  N \mathbf{ i } - \mathbf{Y}$, where $\mathbf{ i } = i \mathbf{ 1 }$,  we recover the  SIS equation, Eq. (\ref{eq:SIS}), after component-wise summation of Eq. (\ref{eq:vectorlogisticequationbycomp}), and division by $N$.
%\begin{eqnarray}
%\frac{d}{dt} i  &=&    -  \delta \frac{1}{N} \sum_j  \mathbf{Y}   +  \beta  \frac{1}{N} \sum_j  \left( (  \mathbf{1}   %  -  \mathbf{Y}  ) \circ \frac{\mathbf{1}}{N-1} \circ  ( N \mathbf{ i } - \mathbf{Y})\right)_j \nonumber \\
%  &=&    -  \delta i   +  \beta  \frac{1}{N}  \frac{1}{N-1}  \left(  N   i   \sum_j    (  \mathbf{1}     -  \mathbf{Y} % )_j       -
%  \sum_j  \left( (  \mathbf{1}     -  \mathbf{Y}  )  \circ   \mathbf{Y})\right)_j \right)\nonumber \\
%  &=&    -  \delta i   +      \beta        i   (  1     -  i )  
%  +      \beta   \frac{1}{N-1}      i   (  1     -  i ) - \beta  \frac{1}{N} \frac{1}{N-1} 
%  \sum_j  \left( (  \mathbf{1}     -  \mathbf{Y}  )  \circ   \mathbf{Y}) \right)_j  \nonumber \\
 %   &=&    -  \delta i   +      \beta        i   (  1     -  i )  
 % +     o\left(N^{-1}\right). \nonumber 
%\end{eqnarray}
\begin{figure}
	  \begin{minipage}[c]{0.50\textwidth}
         \includegraphics[width=\textwidth]{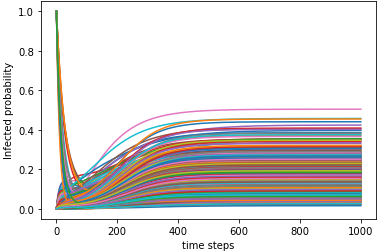}
	    \end{minipage} 
 %\begin{minipage}[c]{0.1\textwidth}	    \end{minipage} 
 	 \begin{minipage}[c]{0.50\textwidth}
	  \includegraphics[width=\textwidth]{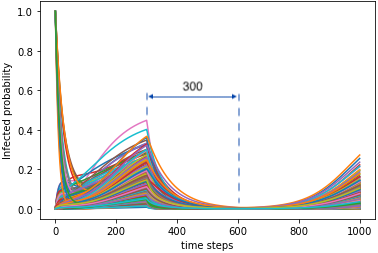}
 	 \end{minipage}
 	 \caption{\scriptsize An illustrative epidemic evolution on a random social contact graph $\mathcal{G}_A(V,E_A)$ with number of nodes N =  1000, number of initially infected nodes $init\_infected = 10$, and average social connectivity  $N\_connect\_A =  40$ (left) .  
% The total number of edges $\vert V_A \vert $ is 49959. 
The variable infection rate  $\beta$ =  0.6(SD0.2), and the  variable recovery rate  $\delta$ =  0.55(SD0.2).  
 % Epidemic fast extinction upper bound, $\beta*spectral\_radius - \delta$ =  0.0999 (endemic).
 % The logistic vector evolution of infected probability, with time step size dt = 0.1, is given for the total number of time steps = 1000.  
 During the confinement, the degree-inducing social contacts  parameter is reduced to $N\_confinement\_A$ =  20. The confinement period, $\Delta t_{conf}$, starts at time 300 and is held on for 300 time steps (right). 
 % The epidemic evolved from average infection probability p\_infect\_0 =  0.01, to p\_infect\_fin = 0.00017159.
 }
 \label{Fig:infectionstatetime}
\end{figure}

\begin{figure}
	  \begin{minipage}[c]{0.475\textwidth}
         \includegraphics[width=\textwidth]{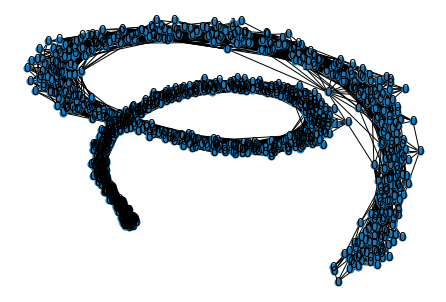}
	    \end{minipage} 
 	     \begin{minipage}[c]{0.05\textwidth}  \hfill	    \end{minipage} 
 	 \begin{minipage}[c]{0.475\textwidth}
	  \includegraphics[width=\textwidth]{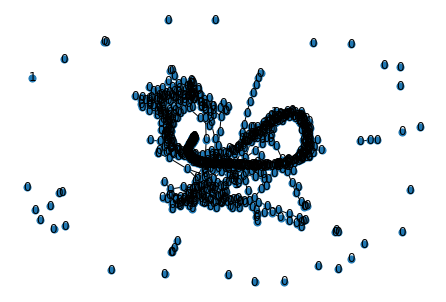}
 	 \end{minipage}
 	 \caption{\scriptsize An unconstrained artificial social contact graph (left) and its confinement rendition (right). In the unconstrained graph ($N=1000, skew = 4, N\_connect\_A = 40$) the mean degree is  25.82, the  mean cluster coefficient is  0.28,  and the graph has 4 connected components.  In the confined graph ($N=1000, skew = 4, N\_confinement\_A = 20$) the mean degree has decreased to 8.04, with 88 connected components resulting. }
 	 \label{Fig:socialgraphs}
\end{figure}

\noindent Finally, with the factor of the social contact graph included in the SIS dynamics, Eq. (\ref{eq:vectorlogistic}), it is now possible to study the effect on the epidemic progression by  changes in the graph structure, see Fig. (\ref{Fig:socialgraphs}). In particular the effect of diminishing the social person-person contacts by culling edges in $\mathcal{G}(V,E)$, see Fig. \ref{Fig:socialgraphs}, while maintaining the degree vector $\mathbf{d}$, allows the dynamical description of confinement efforts in diminishing the epidemic progression.

Concurrently the cost impact from production capacity attrition in the interacting economic graph can be monitored. In the next section, Sec. \ref{sec:economiccapacity}, we define graph-based objective functions for economic capacity. In relation to the social graph, a social cost can be defined proportional to the contact restrictions and the duration of the confinement, see Fig. \ref{Fig:infectionstatetime}, through the quantity $({N_{connect}}_A - {N_{confinement}}_A)\Delta t_{conf}$. The epidemic health cost  can be defined proportional to total infection weight on the social graph at each instance of the epidemic through the quantity $ \int^T\vert \mathbf{Y} (t) \vert_1 dt$.

 \section{The interacting economic capacity network}\label{sec:economiccapacity}
In our present development of the interacting  graphs model we build partially sorted random graphs to resemble real-world configurations - both in social connectivity and economic networks.  In principle  there is no restriction on implementing another topology in either of the interacting graphs. 
\begin{figure}[h]
	  \begin{minipage}[c]{0.50\textwidth}
         \includegraphics[width=\textwidth]{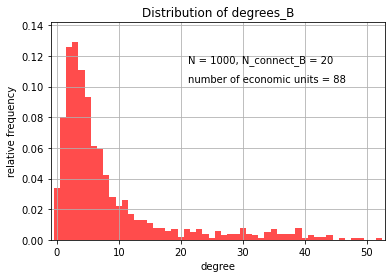}
	    \end{minipage} 
 	     \begin{minipage}[c]{0.1\textwidth}
 
   	    \end{minipage} 
 	 \begin{minipage}[c]{0.50\textwidth}
	  \includegraphics[width=\textwidth]{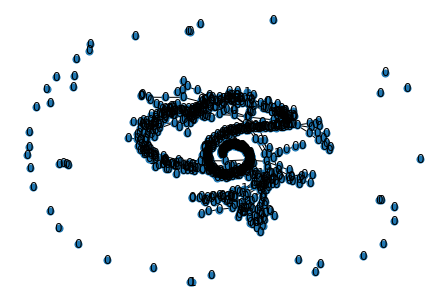}
 	 \end{minipage}
 	 \caption{\scriptsize Construction of the random economic capacity graph $\mathcal{G}_B(V,E_B)$ (right) interacting with the social contact graph $\mathcal{G}_A(V,E_A)$, Fig. \ref{Fig:socialgraphs}.  The economic capacity network inherits the nodes $V_A$ of the social contact graph (N =  1000), and is parametrized by average connectivity parameter  $N\_connect\_B = 20$ to obtain a number of economic capacity units $n_B$ = 88 (left). }
 	 \label{Fig:productiongraph}
\end{figure}
Using the partially sorted random implementation for the economic capacity graph,  the Laplace matrix associated to the adjacency matrix $B$ of $\mathcal{G}_B(V,E_B)$ is used to identify the independent economic units, ${n}_B$ in number. The eigenvectors of the Laplace matrix $\boldsymbol{\delta}_B  \circ \mathbf{1} - B$ with 0-eigenvalue correspond to the connected components of the graph. With ${E_{B}}_i$ the eigenvector of the i-th economic capacity unit, a number of proxy measures for production capacity can be formulated.
 At each instance of time, the  epidemic evolution on the social contact graph $\mathcal{G}_A(V,E_A)$  provides the infected states of all individuals $\mathbf{Y} (t)$.  Using a threshold value $\theta$ in the range $[0,1]$ on the infection probability, the drop-out of active individuals can be assessed at all moment of time.  Then using the \emph{ceiling} function;  $Y_{act}(t) = \vert ceil( \mathbf{Y}(t)  < \theta) \vert_1$, is the number of healthy individuals. Similarly to identify the active individuals of the i-th graph component in optimal situation (no drop-out), we define its binary vector ${\mathbf{W}_B}_i = ceil( {E_B}_i) $.  The epidemic repercussions on each of the economic units can be assessed according the nature of the dependence of the economic output on the active nodes in the economic unit:
\begin{enumerate}
    \item \emph{cumulative metrics} \\
    The drop-out of individuals on the i-th economic component can impact the capacity of the unit proportionally:
   \[ 
   {c_{cum.}}_i =    \mathbf{Y}_{act}(t)^T . {\mathbf{W}_B}_i  
   \]
      The total \emph{cumulative} capacity of the full economic network $\mathcal{G}_B(V,E_B)$ is given by $ {C_{cum}}_B = \sum_{i=1}^{{n}_B}{c_{cum.}}_i $.
    \item  \emph{majority}\\
     The drop-out of individuals on the i-th economic component can impact the integral capacity of the unit by majority support (or other tip-over value):
    \[ 
   {c_{maj}}_i =   ceil\left( \frac{ \vert \mathbf{Y}_{act}(t)^T \circ {\mathbf{W}_B}_i \vert_1}{ \vert {\mathbf{W}_B}_i \vert_1 } \geq .5 \right)  
   \]
   The total \emph{majority} capacity of the full economic network $\mathcal{G}_B(V,E_B)$ is given by $ {C_{maj}}_B = \sum_{i=1}^{{n}_B}{c_{maj.}}_i $.
    \item \emph{fragility} \\
    The drop-out of each single individual of the i-th economic component impacts the integral capacity of the unit:
     \[ 
   {c_{frag}}_i =    \Pi_{j=1}^{\vert {\mathbf{W}_B}_i \vert}  \left( \mathbf{Y}_{act}(t) \cap {\mathbf{W}_B}_i \right)_j
   \]
   where we select by intersection strictly the components corresponding to the i-th economic component, and multiply each.  
   The total \emph{fragile} capacity of the full economic network $\mathcal{G}_B(V,E_B)$ is given by $ {C_{frag}}_B = \sum_{i=1}^{{n}_B}{c_{frag}}_i $.
\end{enumerate}
With the objective functions for economic capacity defined, and a standard expression  for social cost of confinement proportional to  $({N_{connect}}_A - {N_{confinement}}_A)\Delta t_{conf}$ and a health cost proportional to  $ \int^T\vert \mathbf{Y} (t) \vert_1 dt$,   an optimization procedure based on parameters ${N_{confinement}}_A$ and $\Delta t_{conf}$ can be developed. 
\section{Implementation and simulation results}
 A partially sorted random-based social contact graph was implemented to reflect more realistic aspects of true person-person networks as reconstructed by e.g. Barrett et al. \cite{Barrett2009}. In particular the adjacency matrix, $A$, of a graph on N=1000 nodes was designed and parametrized ($skew$, $N\_connect\_A$) to qualitatively approximate the degrees distribution, cluster coefficient distribution and template graph distribution in the communities of Los Angeles, New York City and Seattle \cite{Barrett2009}.  The upper triangular matrix (diag=+1) of an ascending in-row sorted random $N\times N$ matrix in the range $[0,1]$ was used to construct a symmetric matrix $A\_sorted$ with the max values in the upper triangle aligning the main 0-diagonal. Its unsorted counterpart $A\_base$ was retro-fitted by shuffling the row entries right of the main diagonal and restoring symmetry by fitting the lower triangle with the transposed upper triangle matrix. Clearly the sorted proto-adjacency matrix $A\_sorted$ (still with scalars in the range  $[0,1]$) amasses  long linkage and fosters clique formation along the diagonal. In order to tweak this architecture, a parameter $skew$ was used to gradually mix in the sorting effect on the random graph. Finally a degree-indicative connectivity parameter, $N\_connect\_A$, was used to fix the threshold $(N-N\_connect\_A)/N$ for binary adjacency in $A$:\\
 
 $A\_fin = A\_sorted + (A\_base-A\_sorted)/skew$ \hfill \\

 $A = \left(A\_fin >=(N-N\_connect\_A)/N\right)\times 1 $ \hfill\\

A number of parameter configurations where repeatedly tested to show for N = 1000 that $skew = 4$ and $N\_connect\_A = 40 $ lead to an average degree of approximately 26 and an average cluster coefficient of .27 approximately, and qualitatively approximates the degrees distribution and cluster coefficient distribution of true social contact graphs \cite{Barrett2009}. The cluster coefficient distribution can be easily obtained from the adjacency matrix. It is given by the number of unique triangular walks from node $\nu_i$ over the number of contacts in the neighbour sub-graph had it  formed a clique: $\mathit{cc} (\nu_i) = \frac{A^3_{ii} / 2}{ {d_i \choose 2}}$. This sorting and tweaking procedure to construct the artificial social contact network moreover produces cycle and clique template graphs of low degrees. E.g., the particular graph $\mathscr{G}_A(V,E_A)$ in Fig. \ref{Fig:socialgraphs}, counts  111216 of 3-cliques, and  1082464  of 4-cliques. The number of 3-cycli is of course the same as the number 3-cliques, and the number of 4-cycli is  4831030.
\begin{figure}
	  \begin{minipage}[c]{0.50\textwidth}
         \includegraphics[width=\textwidth]{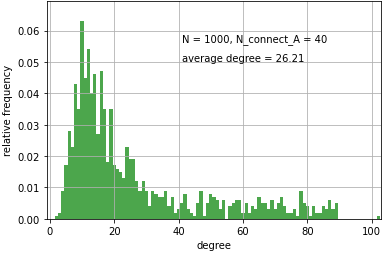}
	    \end{minipage} 
 %\begin{minipage}[c]{0.1\textwidth}	    \end{minipage} 
 	 \begin{minipage}[c]{0.48\textwidth}
	  \includegraphics[width=\textwidth]{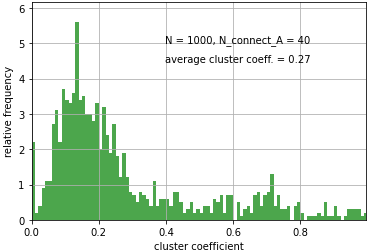}
 	 \end{minipage}
 	 \caption{\scriptsize  The degree distribution of the random artificial social contact graph $\mathscr{G}_A(V,E_A)$ (left), and the corresponding cluster coefficient distribution (right).  }
 \label{Fig:socialgraphditributions}
\end{figure}

\begin{figure}
	  \begin{minipage}[c]{0.50\textwidth}
         \includegraphics[width=\textwidth]{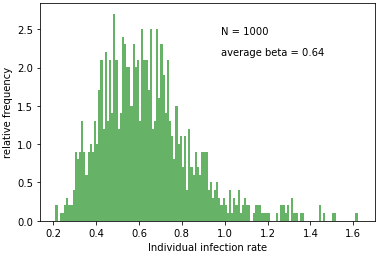}
	    \end{minipage} 
 %\begin{minipage}[c]{0.1\textwidth}	    \end{minipage} 
 	 \begin{minipage}[c]{0.48\textwidth}
	  \includegraphics[width=\textwidth]{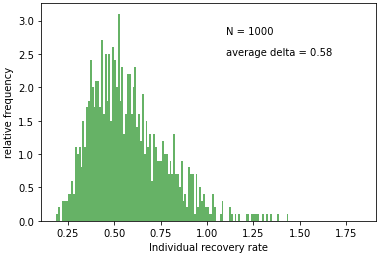}
 	 \end{minipage}
 	 \caption{\scriptsize The lognormal random distributions of infection rate (left) and recovery rate (right) in the random artificial social contact graph $\mathscr{G}_A(V,E_A)$. }
 \label{Fig:betadelta}
\end{figure}

The node-based perspective of the SIS-dynamics was further deployed to randomly attribute individual infection and recovery rates along a lognormal distribution. In particular for the graph $\mathscr{G}_A(V,E_A)$ in Fig. \ref{Fig:socialgraphs}, with infection rate  $\beta$ of  mean  log(0.6) and 0.2 standard deviation and, recovery rate  $\delta$ of  mean  log(0.55) and 0.2 standard deviation, see Fig. \ref{Fig:betadelta}. 
%betaVec = np.random.lognormal( mean=np.log(beta), sigma=np.sqrt(np.log(1+ betaSD**2/beta**2)), size=(N,1))
%deltaVec = np.random.lognormal( mean=np.log(delta), sigma=np.sqrt(np.log(1+ deltaSD**2/delta**2)), size=(N,1))

\noindent We reckon that the social planner impacts  the social contact graph during the confinement period by bringing about the  degree-indicative connectivity parameter to a smaller value $N\_confinement\_A$. This parameter most closely reflects the restriction on the number of  contacts that are allowed in a personal social bubble during confinement.

Finally the multiplicative structure of the right-hand side of the logistic vector first-order differential equation, Eq. \ref{eq:vectorlogistic}, 
\begin{eqnarray}
\dot \mathbf{Y} &=& \mathbf{f}(\mathbf{Y}) \mathbf{Y} \nonumber
\end{eqnarray}
allows a standard solution approach.
From the initial state $\mathbf{Y} (0) = \mathbf{Y}_0$ an incrementally updated solution is obtained. The solutions for the full time-range are obtained by the iterated multiplication with a propagator kernel adapted to the previous state;
\begin{eqnarray}
\mathbf{Y}_{t+1} &=& \left(\mathbf{1}+ \mathbf{f}(\mathbf{Y}_t) dt \right) \mathbf{Y}_t. \nonumber
\end{eqnarray}
In the illustrative epidemic evolution on $\mathcal{G}_A(V,E_A)$, Fig. \ref{Fig:infectionstatetime}, the initial state was randomly seeded with $init\_infected =  10$ nodes. The  evolution of the state vector $\mathbf{Y}$ over the full time range was obtained using time increment $dt = 0.1$ for a total number of time $steps = 1000$.  At the start of the confinement period the propagator kernel is adapted to the reduced adjacency $A_{confined}$ - with original degrees retained -  and applied for next 300 time steps. After the confinement period the original multiplicative kernel based on $A$ is resumed.

\section{Discussion and conclusion}
We explored the possibilities of the vector logistic equation on a social contact graph for the description of contagious disease progression and the description of the possible economic impact of sanitary  measures of  contact regulation.
Our main effort focused on the framing of a graph-based probabilistic SIS-dynamical approach through the vector logistic equation, and constructing interactions with an economic capacity graph. A method of partial-sorting based method was found to implement a social contact graph which resembles more closely some properties of true person-person contact graphs.\\
In real world scenarios the property of social contact is graded. In our present approach the individual's binary adjacencies $A_{ij}$ are only attenuated by the neighbour's proper infection rate $\beta_j$. More realistically this term should include a parameter to express the contact 
\emph{intensity} dependent on each respective contact, i.e. by an infection matrix $\beta_{ij}$ (e.g. related to the time of mutual exposure).\\
Future developments of the graph-based vector logistic dynamics for disease spreading and its economic impact will include development of optimal operational control measures for cost and, refinement of the contamination structure. 

\subsection*{Acknowledgements}
The authors thank the anonymous referees for suggestions on the reciprocal interaction of the economic graph into the social graph. 
 \bibliography{SIS_on_Graph_Arxiv.bib}

\begin{thebibliography}{10}
\providecommand{\url}[1]{\texttt{#1}}
\providecommand{\urlprefix}{URL }
\providecommand{\doi}[1]{https://doi.org/#1}

\bibitem{Allen2007}
Allen, E.: Modeling with It\"o Stochastic Differential Equations.
  Springer-Verlag The Netherlands (2007)

\bibitem{Barrett2009}
Barrett, C.L., Beckman, R.J., Khan, M., Kumar, V.S.A., Marathe, M.V., Stretz,
  P.E., Dutta, T., Lewis, B.: Generation and analysis of large synthetic social
  contact networks. In: Proceedings of the 2009 Winter Simulation Conference
  (WSC). pp. 1003--1014 (2009). \doi{10.1109/WSC.2009.5429425}

\bibitem{Broekaert2020}
Broekaert, J., Busemeyer, J., Pothos, E.: The disjunction effect in two-stage
  simulated gambles. an experimental study and comparison of a heuristic
  logistic, markov and quantum-like model. Cognitive Psychology  \textbf{117},
  101262 (2020). \doi{https://doi.org/10.1016/j.cogpsych.2019.101262},
  \url{https://www.sciencedirect.com/science/article/pii/S001002851930252X}

\bibitem{Buldyrev2010}
Buldyrev, S., Parshani, R., Paul, G., Stanley, H., Havlin, S.: Catastrophic
  cascade of failures in interdependent networks. Nature  \textbf{464},
  1025–1028 (2010). \doi{10.1038/nature08932}

\bibitem{BusemeyerBruza2012}
Busemeyer, J., Bruza, P.: Quantum models of cognition and decision. Cambridge,
  UK: Cambridge University Press (2012)

\bibitem{Choisy2007}
Choisy, M., Gu\'egan, J.F., Rohani, P.: Mathematical modeling of infectious
  diseases dynamics. Encyclopedia of Infectious Diseases: Modern Methodologies
  (2007)

\bibitem{GaneshEtAl2005}
Ganesh, A., Massoulie, L., Towsley, D.: The effect of network topology on the
  spread of epidemics. In: Proceedings IEEE 24th Annual Joint Conference of the
  IEEE Computer and Communications Societies. vol.~2, pp. 1455--1466 vol. 2
  (2005). \doi{10.1109/INFCOM.2005.1498374}

\bibitem{Gray2011}
Gray, A., Greenhalch, D., Hu, L., Mao, X., Pan, J.: A stochastic differential
  equation sis epidemic model. SIAM Journal on Applied Mathematics
  \textbf{71}(3),  876--902 (2011)

\bibitem{Hethcote2000}
Hethcote, H.: The mathematics of infectious diseases. SIAM Review
  \textbf{42}(4),  599--653 (2000). \doi{10.1137/S0036144500371907}

\bibitem{Keeling2005}
Keeling, M.J., Eames, K.T.: Networks and epidemic models. Journal of The Royal
  Society Interface  \textbf{2}(4),  295--307 (2005).
  \doi{10.1098/rsif.2005.0051}

\bibitem{KermackMcKendrick1927}
Kermack, W., McKendrick, A.: A contribution to the mathematical theory of
  epidemics. Proceedings of the Royal Society of London series A
  \textbf{115}(772),  700--721 (1927)

\bibitem{LuEtAl2017}
Lu, D., Yang, S., Zhang, J., Wang, H., Li, D.: Resilience of epidemics for sis
  model on networks. Chaos  \textbf{27}(083105) (2017). \doi{10.1063/1.4997177}

\bibitem{McCormack2006}
McCormack, R., Allen, L.: Stochastic sis and sir multihost epidemic models.
  Proceedings of the Conference on Differential and Difference Equations and
  Applications pp. 775--78 (2006)

\bibitem{Newman2002}
Newman, M.E.J.: Spread of epidemic disease on networks. Phys. Rev. E
  \textbf{66},  016128 (Jul 2002). \doi{10.1103/PhysRevE.66.016128}

\bibitem{Ohtsuki2006}
Ohtsuki, H., Nowak, M.A.: The replicator equation on graphs. Journal of
  Theoretical Biology  \textbf{243}(1),  86--97 (2006).
  \doi{https://doi.org/10.1016/j.jtbi.2006.06.004},
  \url{https://www.sciencedirect.com/science/article/pii/S0022519306002426}

\bibitem{Zhou2006}
Tao, Z., Zhongqian, F., Binghong, W.: Epidemic dynamics on complex networks.
  Progress in Natural Science  \textbf{16}(5),  452--457 (2006).
  \doi{10.1080/10020070612330019}

\bibitem{WangEtAl2003}
Wang, Y., Chakrabarti, D., Wang, C., Faloutsos, C.: Epidemic spreading in real
  networks: an eigenvalue viewpoint. In: 22nd International Symposium on
  Reliable Distributed Systems, 2003. Proceedings. pp. 25--34 (2003).
  \doi{10.1109/RELDIS.2003.1238052}

\bibitem{WangEtAl2013}
Wang, Z., Busemeyer, J., Atmanspacher, H., Pothos, E.: The potential of using
  quantum theory to build models of cognition. Topics in Cognitive Science
  \textbf{5},  672--688 (2013). \doi{10.1111/tops.12043}

\end{thebibliography}
\end{document}